\documentclass[aps,prl,onecolumn,superscriptaddress,nofootinbib,showpacs]{revtex4}
\usepackage{epsfig,amsfonts,amsthm}
\usepackage{amsmath,amssymb}
\usepackage{color}
\newcommand{\be}{\begin{equation}}
\newcommand{\ee}{\end{equation}}
\newcommand{\bea}{\begin{eqnarray}}
\newcommand{\eea}{\end{eqnarray}}

\newcommand{\doublet}[2]{ \left( \begin{array}{c}#1 \\ #2 \end{array}\right) }

\newcommand{\Z}{\mathbb{Z}}
\newcommand{\mmatrix}[4]{ \left(\! \begin{array}{ccc}#1 & #2 \\ #3 & #4 \end{array}\!\right) }

\newcommand{\toCP}{\xrightarrow{CP}}

\def\lsim{\mathrel{\rlap{\lower4pt\hbox{\hskip1pt$\sim$}}
    \raise1pt\hbox{`$<$}}}         
\def\gsim{\mathrel{\rlap{\lower4pt\hbox{\hskip1pt$\sim$}}
    \raise1pt\hbox{$>$}}}         
\newcommand{\lrpartial}{\partial^{\hspace{-6pt}\raise3pt\hbox{\small $\leftrightarrow$}}}

\begin{document}
\title{
{\normalsize \hfill CFTP/16-001} \\*[7mm]
A $CP$-conserving multi-Higgs model without real basis}

\author{I. P. Ivanov}\thanks{E-mail: igor.ivanov@tecnico.ulisboa.pt}
\affiliation{CFTP, Departamento de F\'{\i}sica,
Instituto Superior T\'{e}cnico, Universidade de Lisboa,
Avenida Rovisco Pais 1, 1049 Lisboa, Portugal}
\author{Jo\~{a}o P.\ Silva}\thanks{E-mail: jpsilva@cftp.ist.utl.pt}
\affiliation{CFTP, Departamento de F\'{\i}sica,
Instituto Superior T\'{e}cnico, Universidade de Lisboa,
Avenida Rovisco Pais 1, 1049 Lisboa, Portugal}

\begin{abstract}
Models beyond the Standard Model (bSM) often involve elaborate Higgs sectors,
which can be a source of $CP$-violation.
It brings up the question of recognizing in an efficient way whether a model is $CP$-violating.
There is a diffuse belief that the issue of explicit $CP$ invariance
can be linked to the existence of a basis in which all coefficients are real;
with even a theorem proposed a decade ago claiming
that the scalar sector of any multi-Higgs doublet model
is explicitly $CP$-conserving if and only if
all of its coefficients can be made real by a basis change.
This is compounded by the fact that
in all specific multi Higgs models considered so far, the calculations complied with this claim.
Here, we present the first counterexample to this statement:
a $CP$-conserving three-Higgs-doublet model for which no real basis exists.
We outline the phenomenological consequences of this model,
and notice that the extra neutral Higgs bosons are neither $CP$-even nor $CP$-odd
but are ``half-odd'' under the generalized $CP$-symmetry of the model.
\end{abstract}

\pacs{11.30.Er, 12.60.Fr, 14.80.Ec}

\maketitle

\section{Introduction}

The Standard model uses the minimal Brout-Englert-Higgs mechanism \cite{BEH}.
It breaks the electroweak symmetry, provides masses to the $W$ and $Z$ bosons and to fundamental fermions,
but does not explain the fermion masses and mixing patterns. In particular, the $CP$-violation (CPV) \cite{book}
must be added explicitly in the fermion-Higgs interactions.
On the other hand, it has been known for decades that extended scalar sectors
such as the $N$-Higgs-doublet models (NHDM) can by themselves be a source of $CP$-violation.
Famous classical examples include T.~D.~Lee's 2HDM with spontaneous $CP$-violation \cite{TDLee}
and Weinberg's 3HDM with explicit \cite{Weinberg} or spontaneous \cite{Branco} CPV.
This feature is used nowadays in many bSM models based on non-minimal Higgs sectors.

Models with many fields with the same quantum numbers, including NHDMs, are notorious
for they vast parameter space and reparametrization freedom.
A model can be formulated in a basis which obscures its $CP$ properties.
One way to solve this problem consists in devising basis invariant quantities
which signal $CP$-violation.
Although general methods exist to build such invariants,
the identification of the exact number of necessary and sufficient conditions for
$CP$-conservation must be ascertained on a model-by-model basis
\cite{Gunion-Haber,Gronau:1986xb,Bernabeu:1986fc,Mendez-Pomarol,invariants-J, BotellaSilva, invariants-I, Nishi-CP, Ivanov, Maniatis2008}.
Certain authors seem to believe that
the issue of explicit $CP$ invariance
can be linked to the existence of a basis in which all coefficients are real.
Although the task of finding this real basis can be difficult, its mere {\em existence}
was claimed in \cite{Gunion-Haber} to constitute the necessary and sufficient condition
for explicit $CP$-conservation in the scalar sector. This claim was formulated as Theorem 1
and was proved for generic NHDM with any number of doublets.

In this Letter, we give the first counterexample to this claim.
We construct a model with three Higgs doublets
which is $CP$-conserving, but the coefficients in its scalar potential cannot be made simultaneously real by any basis change.
The explicit $CP$-conservation follows from a generalized $CP$ (gCP) symmetry $J$ of the model,
which maps doublets according to
\be
J:\  \phi_i \toCP X_{ij}\phi_j^*, \quad i = 1,\,2,\,3,\quad \mbox{where}\quad X_{ij} \in U(3)\,,\label{J}
\ee
and which is, for this model, an order-4 transformation.
This symmetry remains intact after EWSB, thus precluding also spontaneous $CP$-violation.
Expressing the lagrangian in terms of physical fields, we find that the extra neutral Higgs bosons
possess a peculiar property: they are neither $CP$-even nor $CP$-odd
but, in a well-defined mathematical sense, are ``half-odd''.

\section{A 3HDM with order-4 gCP}

\subsection{Lagrangian and symmetries}

Consider a 3HDM model with the potential $V = V_0 + V_1$, where
\bea
V_0 &=& - m_{11}^2 (\phi_1^\dagger \phi_1) - m_{22}^2 (\phi_2^\dagger \phi_2 + \phi_3^\dagger \phi_3)
+ \lambda_1 (\phi_1^\dagger \phi_1)^2 + \lambda_2 \left[(\phi_2^\dagger \phi_2)^2 + (\phi_3^\dagger \phi_3)^2\right]
\nonumber\\
&+& \lambda_3 (\phi_1^\dagger \phi_1) (\phi_2^\dagger \phi_2 + \phi_3^\dagger \phi_3)
+ \lambda'_3 (\phi_2^\dagger \phi_2) (\phi_3^\dagger \phi_3)\nonumber\\
&+& \lambda_4 \left[(\phi_1^\dagger \phi_2)(\phi_2^\dagger \phi_1) + (\phi_1^\dagger \phi_3)(\phi_3^\dagger \phi_1)\right]
+ \lambda'_4 (\phi_2^\dagger \phi_3)(\phi_3^\dagger \phi_2)\,,\label{V0}
\eea
with all parameters real, and
\be
V_1 = \lambda_5 (\phi_3^\dagger\phi_1)(\phi_2^\dagger\phi_1)
+ {\lambda_6 \over 2}\left[(\phi_2^\dagger\phi_1)^2 - (\phi_1^\dagger\phi_3)^2\right] +
\lambda_8(\phi_2^\dagger \phi_3)^2 + \lambda_9(\phi_2^\dagger\phi_3)(\phi_2^\dagger\phi_2-\phi_3^\dagger\phi_3) + h.c.
\label{V1b}
\ee
with real $\lambda_5$, $\lambda_6$, and complex $\lambda_8$, $\lambda_9$.
This potential is invariant under the generalized $CP$ transformation (\ref{J}) with
\be
X =  \left(\begin{array}{ccc}
1 & 0 & 0 \\
0 & 0 & i  \\
0 & -i & 0
\end{array}\right)\,.
\label{Jb}
\ee
A key observation is that $J$ is an order-4 transformation:
\be
J^2 = X X^* = \mathrm{diag}(1,\,-1,\,-1) \not = \mathbb{I}\,, \quad J^4 = \mathbb{I} \equiv \mathrm{diag}(1,\,1,\,1) \,.
\ee
For generic values of the coefficients, this potential has no other Higgs-family or gCP symmetries
apart from powers of $J$ \cite{abelian}.
Eqs.~(\ref{V0}) and (\ref{V1b}) define the most general renormalizable potential to which one arrives
starting from any 3HDM invariant under an order-4 gCP and applying basis change transformations
to reduce the number of complex coefficients.

According to the general criterion, the existence of a gCP symmetry indicates that the potential
is explicitly $CP$-conserving \cite{book}. On the other hand, we still have two parameters, $\lambda_8$ and $\lambda_9$,
which remain complex.
There exists no Higgs-basis change that could make all coefficients real.
The proof is elementary.
If such a transformation were possible, then the resulting potential would have
the usual $CP$ symmetry, which is of order 2. Mapping it back to the original basis, we would be able
to find a gCP symmetry of order 2.
But this model does not possess any order-2 gCP symmetry \cite{abelian}.
Therefore, such basis change does not exist.

Thus, the above model is explicitly $CP$ conserving but does not have a real basis, in contradiction with
theorem 1 of \cite{Gunion-Haber}.
This phenomenon requires at least three doublets and is impossible in 2HDM\footnote{Within the 2HDM,
one can impose a gCP symmetry of higher order, dubbed  CP2 and CP3 in \cite{2HDM-GCP}.
However one then arrives at models which possess extra symmetries including the usual $CP$ symmetry.
The model we consider here does not have any other symmetry apart from powers of $J$, see \cite{abelian}.}.
Within the 3HDM, it is the only example of an explicitly $CP$-conserving model without the real basis.
All other symmetry-based 3HDMs were classified in \cite{abelian,3HDM-nonabelian},
and they are either explicitly $CP$-violating or possess an order-2 gCP, which implies existence of the real basis.

The symmetry under $J$ can be extended to the entire lagrangian including the fermion sector.
The simplest choice is to assume that $\phi_2$ and $\phi_3$ are inert doublets which decouple from the fermions altogether.
One can easily arrange for a minimum in which only the first doublet gets a non-zero vacuum expectation value,
making this model a version of the inert doublet model (IDM) with an elaborate inert sector.
In particular, the parameters $\lambda_{3,4}$ and $\lambda'_{3,4}$ in (\ref{V0}) play the role of the IDM's $\lambda_{3,4}$,
the parameters $\lambda_{5,6}$ in (\ref{V1b}) play the role of the IDM's $\lambda_5$
(and, similarly to the IDM, they can be made real),
while the complex parameters $\lambda_{8,9}$ are new and do not have their counterparts in the IDM.

\subsection{Physical scalars and their interactions}

In order to outline the main phenomenological features of the model,
we assume that the second and third doublets are indeed inert and
expand the doublets around the vacuum point $v_1 = v,\, v_2 = v_3 = 0$:
\be
\phi_1 = \doublet{G^+}{{1 \over \sqrt{2}}(v + h_{SM} + i G^0)}\,,\quad
\phi_{2,3} = \doublet{H^+_{2,3}}{{1 \over \sqrt{2}}(h_{2,3} + i a_{2,3})}\,.
\ee
Sorting out terms quadratic in fields, we find that the spectrum of physical scalars is identical to the IDM.
One gets the SM-like Higgs with mass
$m_{h_{SM}}^2 = 2\lambda_1 v^2 = 2m_{11}^2$,
and a pair of degenerate charged Higgses with the mass
$m_{H^\pm}^2 = \lambda_3 v^2/2 - m_{22}^2$.
In the neutral scalar sector, the mass matrices for $h$'s and $a$'s split,
\bea
&&M_{h_2,\,h_3} = \mmatrix{a + b}{c}{c}{a - b}\,,\quad
M_{a_2,\,a_3} = \mmatrix{a - b}{-c}{-c}{a + b}\,,\nonumber\\
&&a = {1 \over 2}v^2 (\lambda_3+\lambda_4) - m_{22}^2 = m_{H^\pm}^2 + {1\over 2}v^2\lambda_4\,,
\quad b = {1 \over 2}v^2\lambda_6\,,\quad c = {1\over 2}v^2\lambda_5\,,
\eea
and lead to the same physical scalar spectrum in both spaces:
\be
M^2 = a + \sqrt{b^2 + c^2}\,,\quad m^2 = a - \sqrt{b^2 + c^2}\,.
\ee
The diagonalization of both mass matrices is performed by a rotation with the angle $\alpha$ defined as
$\tan2\alpha = \lambda_5/\lambda_6$,
but it proceeds in the opposite directions for $h$'s and $a$'s.
The two heavier scalars $H, A$ and the two ligher scalars $h$ and $a$ are related to initial fields as
\be
\doublet{H}{h} = \mmatrix{c_\alpha}{s_\alpha}{-s_\alpha}{c_\alpha}\doublet{h_2}{h_3}\,,\quad
\doublet{a}{A} = \mmatrix{c_\alpha}{s_\alpha}{-s_\alpha}{c_\alpha}\doublet{a_2}{a_3}\,.\label{rotations}
\ee
Applying this rotation to the doublets $\phi_2$ and $\phi_3$ leads to
\be
c_\alpha \phi_2^0 + s_\alpha\phi_3^0 = {1 \over\sqrt{2}}(H + i a)\,,
\quad
-s_\alpha \phi_2^0 + c_\alpha\phi_3^0 = {1 \over\sqrt{2}}(h + i A)\,.
\ee
Thus, judging only by the scalar spectrum, this model closely resembles a duplicated
IDM with two mass-degenerate copies of the inert sector.
We also stress that the mass degeneracy between neutral scalars arises in this model not because of an extra $U(1)$
symmetry but as a result of imposing a single discrete gCP of order 4.

\subsection{$CP$-eigenstates}

The fact that the neutral Higgs spectrum contains two pairs of mass degenerate fields
allows one to redefine the degrees of freedom in a suitable way.
The real neutral fields $H, A, h, a$ are not eigenstates of the gCP symmetry:
\be
J: \quad H \toCP A\,, \quad A\toCP -H\,, \quad h \toCP -a\,, \quad a\toCP h\,.
\ee
But since $J$ is conserved,
it is natural to combine them into {\em complex} neutral fields,
\be
\varPhi = {1\over \sqrt{2}}(H - i A)\,, \quad \varphi = {1\over \sqrt{2}}(h + i a)\,,\label{bizarre}
\ee
which {\em are} the eigenstates of mass and of $CP$:
\be
J: \quad \varPhi \toCP i\varPhi\,,\quad \varphi \toCP i\varphi\,.\label{J-eigenstates}
\ee
One can associate the gCP transformation with the global quantum number $q$, which is defined modulo 4,
and assign $q=+1$ to $\varPhi$, $\varphi$, and $q=-1$ to their conjugate fields.
All other neutral fields are either $CP$-odd, $q=+2$, or $CP$-even, $q=0$.
Since $J$ is a symmetry of the lagrangian and of the vacuum, it commutes with the hamiltonian.
Therefore, in any transition between initial and final states with definite $q$, this quantum number is conserved.

It is instructive to compare the $CP$-properties of the inert neutral Higgses in this model and in the usual IDM.
In both models, we have a gCP symmetry: either the standard $CP$ in the IDM or the gCP symmetry $J$ of order 4.
Both models feature a $\Z_2$ family symmetry associated with the inert scalars.
The difference is that, in this model, this extra symmetry is generated not by a new transformation but by $J^2$.
In the IDM, the mass eigenstates of the neutral inert scalar fields, usually denoted $H$ and $A$,
are associated with $CP$-even and $CP$-odd scalars. In the absence of direct couplings to the fermions,
the assignment which is which is undefined, as it depends on our choice of which of the two gCP symmetries
of the IDM is labeled as ``the $CP$ symmetry''. However, irrespective of the choice, we have two neutral scalars
of opposite $CP$ parity.

In the present model, in contrast to the IDM, the mass and $CP$-eigenstates can be chosen in such as way that
they have the same $CP$-properties, {\em c.f.} (\ref{J-eigenstates}). Again, here we also have two possible choices
for gCP: $J$ or $J^3$. But, the two fields $\varPhi$ and $\varphi$ still have the same $CP$-properties irrespective of this choice.
Most crucially, here, unlike in the IDM, the inert scalars are neither $CP$-even nor $CP$-odd but ``$CP$-half-odd'',
which is reflected by their $q$-charge being one half of the $q$-charge of a $CP$-odd quantity.
One would need a pair of such fields to construct a $CP$-odd object.

Counting the degrees of freedom in the neutral inert sector, we started from two complex fields $(\phi_2^0,\,\phi_3^0)$
with undefined $CP$-properties, and by splitting into four real fields and regrouping them again, arrived at two other
complex fields $\varPhi,\, \varphi$. One can remove the intermediate fields altogether and represent this transformation
with a non-holomorphic norm-preserving linear map in $\mathbb{C}^2$:
\be
\doublet{\varPhi}{\varphi} = \left(\!\begin{array}{cc} c_\gamma & s_\gamma \\ -s_\gamma & c_\gamma \end{array}\!\right)
{1\over \sqrt{2}}\doublet{\phi_2^0 + \phi_3^{0*}}{\phi_3^0 - \phi_2^{0*}}\,,\label{C2map}
\ee
with $\gamma = \alpha - \pi/4$.
Clearly, this transformation goes beyond the simple basis changes as it mixes $\phi$'s
and $\phi^*$.
It is somewhat similar to the transformations studied in \cite{Pilaftsis} within the 2HDM,
but in our case it acts not on doublets but on their neutral components only.

The map (\ref{C2map}) preserves the norm and, therefore, the kinetic term of the inert scalars:
\be
|\partial_\mu\phi_2^0|^2 + |\partial_\mu\phi_3^0|^2 = |\partial_\mu\varPhi|^2 + |\partial_\mu\varphi|^2\,.
\ee
One can also rewrite the other parts of the lagrangian, after EWSB, in terms of these neutral fields.
For example, the interactions of the SM-like Higgs with the inert scalars take a simple form:
\be
\left(2{h_{SM}\over v} + {h_{SM}^2 \over v^2}\right)\left[(m_{H^\pm}^2 + m_{22}^2)(H_2^+H_2^- + H_3^+H_3^-) +
(M^2 + m_{22}^2) \varPhi^*\varPhi + (m^2 + m_{22}^2) \varphi^*\varphi\right]\,.\label{Vhi}
\ee
The gauge field interactions with the inert neutrals contain
\be
{1\over 2}\left(g^2 W_\mu^+W_\mu^- + \bar{g}^2 Z_\mu Z_\mu\right)(\varPhi^*\varPhi + \varphi^*\varphi)
+ i {\bar g\over 2\sqrt{2}}Z_\mu \left(\varphi \lrpartial_\mu \varPhi - \varphi^* \lrpartial_\mu \varPhi^*\right)\,,\label{Zmu}
\ee
with the usual notation $\bar g \equiv \sqrt{g^2 + g^{\prime 2}}$.
The last term induces transitions $Z \to \varPhi\varphi$ or $Z \to \varPhi^* \varphi^*$,
with the two scalars being in the $p$-wave.
Notice that in the IDM, the corresponding transition $Z \to HA$
involves two scalars with opposite $CP$ parities, so that their product, the combined internal $CP$ parity of the pair, is negative:
$(+1)\cdot(-1) = -1$.
Here, the transition involves scalars with the {\em same} $CP$-properties.
The product of internal $CP$ properties of the scalar pair is still $CP$-odd,
but it arises as a combination of two ``$CP$-half-odd'' quantum numbers: $i\cdot i = -1$.

Inert scalars also interact with themselves via the quartic potential.
Written with the new fields, the potential contains terms such as
$\varphi^*\varphi$, $\varphi^4$, $(\varphi^*)^4$, or $\varphi^2(\varphi^*)^2$, where any $\varphi$ can also be replaced by $\varPhi$.
For example, the terms in the potential (\ref{V1b}) with the imaginary parts of $\lambda_8$ and $\lambda_9$
are given by the combination $i(\varphi^2  + \varPhi^2 - \varphi^{*2} - \varPhi^{*2})$
coupled to
\be
\varPhi^*\varphi^* + \varPhi \varphi\,,\quad \varphi^2 + \varphi^{*2} - \varPhi^2 - \varPhi^{*2}\,,
\quad H_2^+H_2^- - H_3^+H_3^-\,.
\ee
All of these interactions conserve the quantum number $q$.
They can induce transitions such as $\varphi^* \to \varphi\varphi\varphi$ or $\varphi\varphi \to \varphi^*\varphi^*$,
and, via loops, mixing between $\varPhi$ and $\varphi$ (which is just a non-diagonal contribution to their renormalization),
but they never transform $\varphi$ into $\varphi^*$.

It is also worth noticing that, when written in terms of complex fields $\varphi$ and $\varPhi$,
the potential does not possess the $U(1)$ symmetry of their arbitrary phase rotations.
Still, their real and imaginary parts are mass degenerate, and it is the residual order-4 rephasing symmetry (\ref{J-eigenstates})
that links them and prevents the mass splitting.

\section{Discussion}

First, let us pinpoint the loophole in the proof of Theorem 1 in \cite{Gunion-Haber} which allows for
the counterexample presented here. The proof assumes that the $T$-transformation operator squares to identity,
with a reference to \cite{phase-factors}. But references \cite{phase-factors} explicitly mention that they restrict themselves to
discrete symmetry transformations accompanied only by phase factors, which, in our notation, correspond to the gCP transformations
(\ref{J}) with diagonal matrices $X$.
However, $X$ can be non-diagonal and can be simplified to a block-diagonal form with $2\times 2$ blocks, \cite{gcp-standard}.
Our $X$ is precisely of this type, and the assumption that $T$-transformation squared is equal to the identity transformation
does not apply anymore.

Our example modifies the status of Theorem 1 of \cite{Gunion-Haber} by limiting its applicability.
The theorem remains valid for explicitly $CP$-conserving models containing an order-2 gCP.
However the absence of a real basis does not necessarily mean that the model is $CP$-violating,
as it might contain a higher-order gCP symmetry. Existence of a real basis is a sufficient but not necessary
condition for a model to be explicitly $CP$-conserving.

One might ask whether similar $CP$-conserving models based on a higher order gCP symmetry are possible.
Here, we have two remarks. First, the order of a gCP transformation $J$ must be a power of 2.
If it were not, it would be possible to define a new gCP transformation $J'$ of smaller order and generate the symmetry group
with $J'$ and a horizontal symmetry transformation.
Second, a model with a higher order $J$ definitely goes beyond three doublets. In particular, it was checked in \cite{abelian}
that trying to define a 3HDM model with an order-8 gCP symmetry leads to a potential with an accidental continuous symmetry.

Next, coming to the phenomenology of the model, a quick check shows that,
despite involving complex parameters in their interaction potential,
the inert scalars do not produce physically detectable signs of $CP$-violation.
First, the new scalars do not directly couple to fermions.
Second, the $Z$-boson coupling to a pair of scalars follows that same pattern as in the $CP$-conserving 2HDM \cite{book,undoubtable}.
Third, although the charged scalars appear in $h_{SM}\gamma\gamma$ or $h_{SM}Z\gamma$ loops,
they do not induce the $CP$-odd $h_{SM}F_{\mu\nu}\tilde F^{\mu\nu}$ effective vertex.

Similarly to the usual inert doublet model, the lightest inert scalar, which we take to be $\varphi$,
is protected against decay and is the dark matter candidate.
The heavier scalar, $\varPhi$, can decay, for example, via $\varPhi \to Z^{(*)}\varphi \to \mathrm{SM}+\varphi$,
induced by the interaction (\ref{Zmu}).

The most intriguing question is whether the remarkably unusual $CP$-properties of the inert scalars
can be at all experimentally detected. Notice that if such an experiment were possible, it would have dramatic consequences,
as it would distinguish this model with ``$CP$-half-odd'' states from any form of the usual inert doublet model
with $CP$-even or $CP$-odd scalars. In other words, such an experiment would distinguish order-2 gCPs from order-4 gCP,
which are so far considered to be experimentally indistinguishable.
At present, we do not have any concrete suggestion of a measurement capable of accomplishing this task.

\section{Conclusions}

In summary, we found a bSM model which dispels the popular belief that a $CP$-conserving
scalar sector necessarily requires existence of a real basis.
In the light of our work, the questions of what are the true necessary and sufficient conditions
for explicit $CP$-conservation and whether they distinguish order-2 from higher order gCP symmetries must be reexamined.

Our example is a three-Higgs-doublet model featuring an order-4 generalized $CP$-symmetry
and no other symmetries.
We found that the phenomenology of this model is similar to the usual inert doublet model
but with inert neutral Higgses possessing the remarkably unusual property of being ``half-odd'' under the $CP$ symmetry of the model.
Whether the exotic property of being $CP$-half-odd is measurable, at least in principle, or if it is just a mathematical peculiarity,
remains an open and a very intriguing question.
\\

We thank Gustavo Branco, Ilya Ginzburg, Howard Haber, Lu\'is Lavoura, and Andreas Trautner for useful comments and discussions.
This work was supported by the Portuguese
\textit{Fun\-da\-\c{c}\~{a}o para a Ci\^{e}ncia e a Tecnologia} (FCT)
under contracts UID/FIS/00777/2013 and CERN/FIS-NUC/0010/2015,
which are partially funded through POCTI (FEDER), COMPETE, QREN, and the EU.
I.P.I. acknowledges funding from the \textit{Funda\c{c}\~{a}o para a Ci\^{e}ncia e Tecnologia}
through the FCT Investigator contract IF/00989/2014/CP1214/CT0004
under the IF2014 Programme.

\end{document}